# Citation

Issa Atoum, Ahmed Otoom and Narayanan Kulathuramaiyer. Article: A Comprehensive Comparative Study of Word and Sentence Similarity Measures. *International Journal of Computer Applications* 135(1):10-17, February 2016. Published by Foundation of Computer Science (FCS), NY, USA.

```
@article{key:article,
        author = {Issa Atoum and Ahmed Otoom and Narayanan Kulathuramaiyer},
        title = {Article: A Comprehensive Comparative Study of Word and Sentence Similarity Measures},
        journal = {International Journal of Computer Applications},
        year = {2016},
        volume = {135},
        number = {1},
        pages = {10-17},
        month = {February},
        note = {Published by Foundation of Computer Science (FCS), NY, USA}
}
```

# A Comprehensive Comparative Study of Word and Sentence Similarity Measures

Issa Atoum
Faculty of Computer Information
The World Islamic Sciences &
Education University
11947 Amman, Jordan
Issa.Atoum@wise.edu.jo

Ahmed Otoom
Royal Jordanian Air forces
11134 Amman, Jordan
aotoom@rjaf.mil.jo

Narayanan Kulathuramaiyer
Faculty of Computer Science and
Information Technology
Universiti Malaysia Sarawak
94300 Kota
Samarahan,Sarawak,Malaysia
nara@fit.unimas.my

## ABSTRACT
Sentence similarity is considered the basis of many natural language tasks such as information retrieval, question answering and text summarization. The semantic meaning between compared text fragments is based on the words' semantic features and their relationships. This article reviews a set of word and sentence similarity measures and compares them on benchmark datasets. On the studied datasets, results showed that hybrid semantic measures perform better than both knowledge and corpus based measures.

## 1. INTRODUCTION
Semantic similarity finds a resemblance between the related textual terms. Words are considered semantically similar or related if they have common relationships. For example, *food* and *salad* are semantically similar; obviously *salad* is a type of *food*. Also, *fork* and *food* are related; undoubtedly a *fork* is used to take *food*. Resnik illustrated that word similarity is a subcase of word relatedness[1].

The word similarity is the foundation of the sentence similarity measures. A Sentence similarity method measures the semantics of group of terms in the text fragments. It has an important role in many applications such as machine translation [2], information retrieval [3]–[5], word sense disambiguation [6], spell checking [7], thesauri generation [8], synonymy detection [9], and question answering [10]. Furthermore, semantic similarity is also used in other domains; in medical domain to extract protein functions from biomedical literature [11] and in software quality[12]–[14] to find common software attributes.

Generally, sentence similarity methods can be classified as corpus based, knowledge based and hybrid methods. Corpus based methods depend on building word frequencies from specific corpus. In this category, Latent Semantic Analysis (LSA) [10], [15], [16], and Latent Dirichlet Allocation (LDA) [3], [17], [18] have shown positive outcomes, however they are rather domain dependent [19], [20]. In other words, if the model (i.e. corpus model) was built for news text, it usually performs poorly in another domain such computer science text.

The knowledge based methods usually employ dictionary information such as path and/or depth lengths between compared words to signify relatedness. These methods suffer from the limited number of general dictionary words that might not suit specific domains. Most knowledge based measures depend on WordNet[21]. WordNet is a hand crafted lexical knowledge of English that contains more than 155,000 words organized into a taxonomic ontology of related terms known as synsets. Each synset (i.e. a concept) is linked to different synsets via a defined relationship between concepts. The most common relationships in WordNet are the 'is-a' and 'part –of' relationships.

Hybrid methods combine the corpus based methods with knowledge based methods and they generally perform better.

To the best of authors knowledge, there are a few works that compares sentences [22] [10]. This article compares state of the art word and sentence measures on benchmark datasets. It is found that hybrid measures are generally better than knowledge and corpus based measures.

## 2. RELATED WORK
### 2.1 Word Similarity Methods
#### 2.1.1 Corpus based Methods
These methods depend on word features extracted from a corpus. The first category of these methods is based on the information content (IC) of the least common subsumer (LCS) of compared term synsets [23]–[25]. The second category, a group known as distributional methods, depends on distribution of words within a text context. Words co-occurrences are represented as vectors of grammatical dependencies. The distributional method, LSA similarity [16], [26] transforms text to low dimensional matrix and it finds the most common words that can appear together in the processed text. Corpus based methods are domain dependent because they are limited to their base corpora.

#### 2.1.2 Knowledge based Methods
Knowledge based methods use information from dictionaries (such as WordNet) to get similarity scores. Classical knowledge based methods use the shortest path measure [27] , while others extend the path measure with depth of the LCS of compared words [28], [29] . Leacock Chodorow [30] proposed a similarity measure based on number of nodes in a taxonomy and shortest path between compared terms. Hirst and St-Onge [31] considered all types of WordNet relations; the path length and its change in direction. Some methods [23]–[25] have the ability to use intrinsic information rather than information content. Knowledge based methods suffer from limited hand crafted ontologies.

### 2.1.3 Hybrid Methods

Hybrid based methods associate functions from corpus and knowledge based methods. Zhou *et al.* [32] proposed a similarity measure as a function of the IC and the path length of compared words. Rodriguez and Egenhofer [33] used the weighted sum between synsets paths, neighboring concepts and their features in a knowledge fusion model. Dong *et al.* [34] proposed a weighted edge approach to give different weights of words that share the same LCS and have the same graph distance; words with lower edge weights are more similar than words with higher edge weights. Atoum and Bong [35] proposed a hybrid measure of distance based/knowledge based method[29] and information content method [23]. They called their model Joint Distance and Information Content Word Similarity Measure (JDIC).

In this category also, web based methods depend on the web resources to calculate the similarity. Turney *et al.* [9] used a measure called Point-Wise Mutual Information and Information Retrieval (PMI-IR) that is based on the number of hits returned by a web search engine. Bollegala *et al.* [36] used a WordNet metric and Support Vector Machines on text snippets returned by a Web search engine to learn semantically related and unrelated words.

## 2.2 Sentence Similarity Methods

### 2.2.1 Corpus based Methods

These methods are based on a corpus features. The first category, traditional information retrieval methods, Term Frequency –Inverse Document Frequency (TF-IDF) methods [37]–[39], assume that documents have common words. However, these methods are not valid for sentences because sentences may have null common words[29], [40] . For example, the sentences *"my boy went to school"* and *"kids learn math"* do not have any common word although they are semantically related to education.

Based on the TF-IDF idea, the second category, word co-occurrence methods are proposed. They model words co-occurrences as vectors of semantic features; LSA[10][16][26], Hyperspace Analogues to Language (HAL) [41], and LDA [7][17][18][45]. After these vectors are processed a similarity measure such as the cosine measure is used to calculate the final similarity between compared text fragments.

The third category, string similarity methods (mini corpus based methods) depend on strings edit distance and the word order in a sentence [43]–[45].The fourth category, the gigantic corpus methods. They use the internet resources as their baseline; Wikipedia [46], Google Tri-grams[6][47] , and search engine documents [48]. These methods are more practical to text rather than sentences.

Corpus based methods (second and fourth category) suffer from these problems; once the vector space model is built for a domain it can be hardly used in another domain [19]. In addition, adding new instance of existing model becomes infeasible, as it requires rebuilding the whole model, (i.e. computationally costly). They also have the problem of high sparse vectors especially for short sentences and generally they are not practical [20].

### 2.2.2 Knowledge based Methods

knowledge based methods use semantic dictionary information such word relationships [31][40][49], information content [1], [23] to get word semantic features. Li *et al.* [20] proposed a sentence similarity based on the aspects that a human interprets sentences; objects the sentence describes, properties of these objects and behaviors of these objects.

Tian *et al.* [19] proposed sentence similarity based on WordNet IC and part of speech tree kernels. Huang and Sheng [45] proposed a sentence similarity measure for paraphrase recognition and text entailment based on WordNet IC and string edit distance. Lee [50] built semantic vectors from WordNet information and part of speech tags. Abdalgader and Skabar [51] proposed a sentence similarity measure based on word sense disambiguation and the WordNet synonym expansion. Tsatsaronis *et al.* [40] measured the semantic relatedness between compared texts based on their implicit semantic links extracted from a thesaurus. Li *et al.* [52] proposed a sentence similarity measure based on word and verb vectors and the words order.

Generally, the knowledge based methods are limited to the human crafted dictionaries. Due to this, not all words are available in the dictionary and even if a few words exits they usually do not have the required semantic information. As an example, WordNet has a limited number of verbs and adverbs synsets compared to the list of available nouns synsets in the same ontology.

### 2.2.3 Hybrid Methods

Hybrid methods are a combinations of the previous mentioned methods. Croft *et al.* [4] applied their measure on photographic description data based semantic vectors of path and term frequency. Li *et al.* [29] proposed a sentence similarity based on WordNet information, IC of Brown Corpus, and sentence words orders. Later, [52] proposed a word similarity based on a new information content formula and Lin word similarity[23].

Ho *et al.* [6] incorporated a modified version of word sense disambiguation of [53] in their similarity measure. Feng *et al.* [54] used direct( words relationships) and indirect (reasoning) relevance between sentences to estimate sentence similarity. Liu *et al.* [55] proposed a sentence similarity based on Dynamic Time Wrapping (DTW) approach. They calculated the similarity between sentences by aligning sentences parts of speech using DTW distance measure. Ho *et al.* [6] showed that DTW is computationally costly and time proportionately with the sentence's length.

A combination of eight knowledge base measures and three corpus based measures is proposed in [39], [56]. The final word similarity measure is the average of all eight measures. The sentence similarity measure is derived using word overlapping over an IDF function of words in related segments.

Hybrid approaches shows promising results on standard benchmark datasets. Table 1 shows the summary of different word and sentence similarity measures.

Table 1. Summary of word and sentence similarity approaches

| Similarity Method | Approach | Advantages | Disadvantages |
|---|---|---|---|
| Corpus based methods | Use a corpus to get probability or frequency of a word in a corpus | Preprocessed corpus to reduce computations | 1. Corpus is domain dependent. 2. Some words might get same similarity. 3. Semantic vectors are sparse. |
| Knowledge based methods | Use dictionary information such as WordNet to get similarity (for example, path and depth, word relationships, etc.) | Adoptions of human crafted ontology can increase accuracy | 1. Limited words. 2. Some words can get same similarity if they have the same path and depth |
| Hybrid methods | Use both corpus and a dictionary information. | Usually performs better | 1. Additional computations |

## 3. EXPERIMENTAL EVALUATION
### 3.1 Word Similarity Methods

To evaluate the performance of word similarity methods, the Rubenstein Goodenough[57] and Miller Charles[58] word pairs benchmark datasets are selected. Rubenstein and Goodenough investigated synonymy judgements of 65 noun pairs categorized by human experts on the scale from 0.0 to 4.0. Miller and Charles selected 30 word pairs out of the 65 pairs of nouns and organized them under three similarity levels.

The experiments were run with WordNet 3.0 [59] for knowledge based measures and Brown Dictionary [60] for corpus based measures. The similarity measures are implemented using python custom code. Figure 1 and Figure 2 respectively summarizes the Pearson correlation of different similarity measures against human means on the Miller and Goodenough datasets.

Results showed it cannot be argued what is the best word method unless the method is used in real application or tested on a benchmark dataset. However, hybrid methods (e.g. JDIC) perform better than other corpus and knowledge based methods.

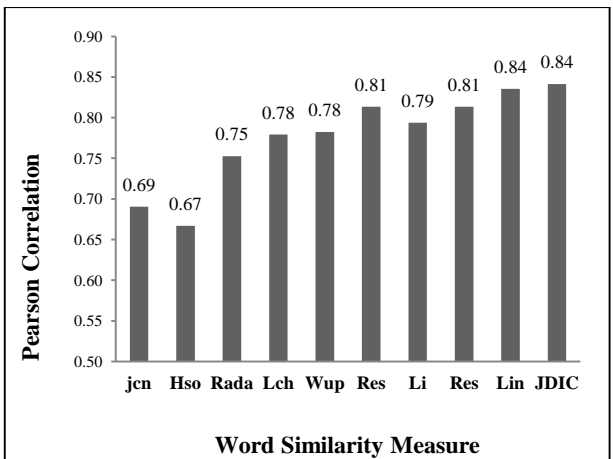

Fig 1: Pearson Correlation versus word similarity measures on Miller dataset

### 3.2 Sentence Similarity Methods
To evaluate the performance of the sentence similarity methods, the dataset constructed by [29] ( the STSS-65 dataset) is selected[1]. It consists of sentences pairs that were originally constructed manually to evaluate a short similarity measure named STASIS. In STSS-65 dataset, the corresponding words in [57] are replaced with the words definitions from the Collins Cobuild Dictionary [61]. Instead of keeping all the 65 pairs Li *et al.* [29] decided to keep only the most accurate annotated and balanced sentence pairs. Note that in this dataset, the pair number 17 has been used with different Human scores namely (0.13,0.063,0.048) in different research works e.g., [4], [29], [50]. The human score 0.13 was first used in the main work of [29], but later [62] published the dataset on 2009 with the figure 0.048 (0.19 from 4). The 0.13 figure is used in this article as first used by the original work of [29].

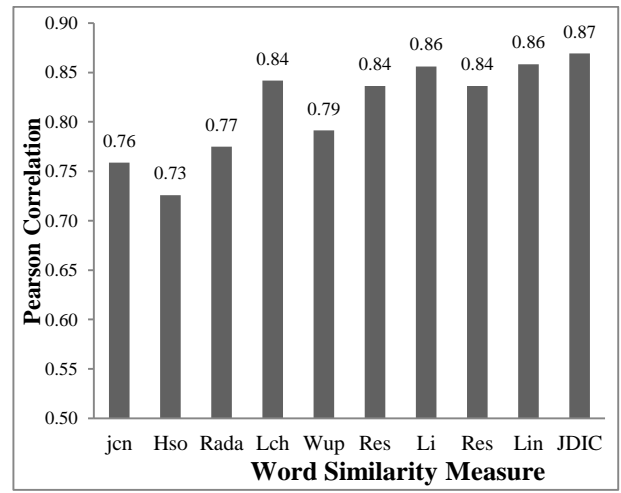

Fig 2: Pearson Correlation versus word similarity measures on Rubenstein Goodenough dataset

For all experiments WordNet Version 3.0 is used. For Mihalcea [11] measure the PMI-IR measure is replaced with Normalized Search engine Index Distance (NSID) [63] as Turner 's PMI is not available. Also, Wikipedia dataset of December 2013 were used for LSA measure and Open America National Corpus (OANC) to replace BNC Corpus.

Table 2 shows the Pearson correlation and Spearman's rank coefficient between different measures and Human participants' ratings. On the first hand, the Pearson correlation is either calculated or taken from respected works. On the other hand, the Spearman's rank figure is calculated using published similarity figures of the respected works. The computed similarity scores are sorted in an ascending order, and the ranking of similarities is compared against the benchmark dataset using Spearman's rank correlation.

---
[1] http://semanticsimilarity.net/benchmark-datasets.

**Table 2. Pearson and Spearman correlations with respect to human ratings on STS-65 dataset**

| Similarity Measure | Pearson Correlation | Spearman Correlation |
|---|---|---|
| Worst Human Participant | 0.590 | N/A |
| Ming Che Lee 2011[50] | 0.705 | 0.661 |
| Mihalcea *et al.* 2009 [39] | 0.708 | 0.687 |
| Feng et al. 2008 [54] | 0.756 | 0.649 |
| Croft *et al.* 2013 (LSS) [4] | 0.807 | 0.810 |
| Li *et al.* 2006(STASIS)[29] | 0.816 | 0.804 |
| Mean of all Human Participants | 0.825 | N/A |
| O'shea *et al.* 2008 (LSA)[10] | 0.838 | 0.811 |
| Liu *et al.* 2007 [55] | 0.841 | 0.853 |
| Islam *et al.* 2008 [43] | 0.853 | 0.828 |
| Tsatsaronis *et al* 2010 (Omiotis)[40] | 0.856 | 0.890 |
| Ho *et al.* 2010 (SPD-STS)[6] | 0.895 | 0.905 |
| Islam *et al.* 2012(Tri-Grams) [47] | 0.914 | 0.798 |

Table 2 shows that Ming [50] and Mihalcea measures have the lowest Pearson and Spearman Coefficients. To investigate this result, Mihalcea [11] is taken as an example. Each of the 8 different measures (of Mihalcea) has its strengths and weakness. One of them, Wikipedia measure has relatively high similarity (>0.5) while the path measure has relatively low similarity (<0.1). Therefore, once the average all the measures is computed the final similarity score will be no longer be near the human similarity rating score. More precisely, from Mihalcea's study got score values in range (0.07-0.5) for all compared benchmark sentence pairs. The authors findings resemble Ho *et al.* [6] findings. They showed that simple average similarity can never be a good similarity measure.

Many sentence similarity approaches have been proposed but many of them might be difficult to implement[47], [64] or has poor performance[4], [50], [54], [64] . For example the works of [39], [56] are based on 8 different knowledge based measures and 3 corpus based measures which makes their implementation difficult. Further difficulties in other works includes the need of processing gigantic data processing [47].

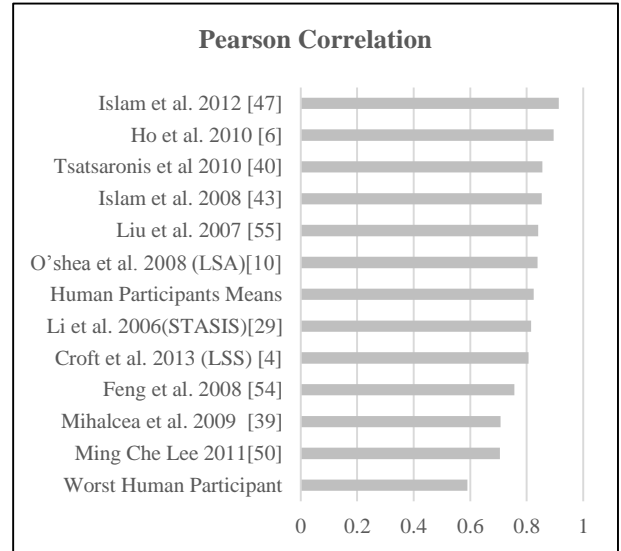

**Fig 3: Pearson correlation versus sentence similarity measures on STS-65 dataset**

[47] used the Web 1T 5-gram dataset; a compressed text file of approximately 24 GB compressed composed of more than 1 million tri-grams extracted from 1 trillion tokens. Nevertheless, [47] [10] are considered comprehensive datasets and can be accessed easily once indexed.

Figure 3 shows the similarity measure versus Pearson Correlation over the STS-65 dataset. Table 2 shows that hybrid methods (e.g. [6], [55], [40]*)* perform better than knowledge based (e.g. [29]) and corpus based (e.g. [10]) methods. Islam *et al.* Tri-gram measure [47] is an exception. This finding is explained by studying details in Table 3. Table 3 shows the STS-65 benchmark dataset word pairs (second column) that correspond to the list of sentences (i.e. sentences used in similarity measures). The human mean score rating (third column), in the range of 0.01-0.96, represent dissimilar to very similar sentences. It is found that [47] overestimates the human rating scores especially the dissimilar sentence pairs. Conversely, this finding was not clear at the Pearson correlation level shown in Table 2.

Figure 4 shows the STS-65 dataset human scores versus the scores of [47] and [40]. It is clear that [47] overestimates sentence pairs 1-29(30% of the original dataset). However, the same method works well for pairs that are semantically similar as per human scores (30-65). On the other hand, although [40] has less Pearson correlation, as shown in Figure 3, it is relatively better than [47] in sentence pairs 1-29. Therefore, the Pearson correlation (in this case) is not a good measure to compare sentence measures that are relatively dissimilar. It is concluded that another measure should take into consideration this case instead of using an average as in the case of Pearson correlation.

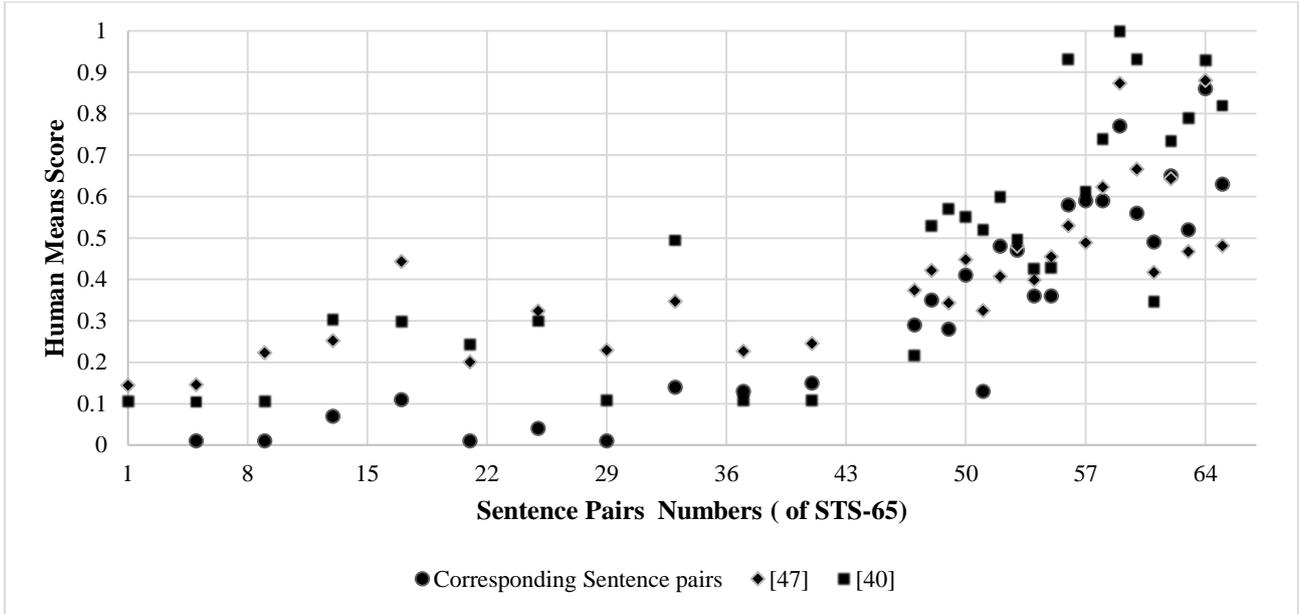

**Fig 4: Pearson correlation versus sentence similarity measures on STS-65 dataset**

## 4. CONCLUSION

This article studies a set of word and sentence similarity measures. The study showed that word similarity is not enough to select a *good* sentence similarity measure. Hybrid sentence methods are generally better than corpus and knowledge based methods. In the future, it is planned to test more word and sentence methods on other datasets. Furthermore, more work will concentrate on an approach to choose between Spearman and Pearson correlations.

**Table 3. STS-65 dataset results**

| No | Corresponding word pairs | Mean | Li 2006 | Tsatsaronis 2010 | Islam 2008 | Ho 2010 | Croft 2013 | Islam 2012 |
|---|---|---|---|---|---|---|---|---|
| 1 | cord-smile | 0.01 | 0.33 | 0.11 | 0.06 | 0.09 | 0.18 | 0.14 |
| 5 | autograph-shores | 0.01 | 0.29 | 0.10 | 0.11 | 0.06 | 0.20 | 0.15 |
| 9 | asylum-fruit | 0.01 | 0.21 | 0.10 | 0.07 | 0.04 | 0.28 | 0.22 |
| 13 | boy-rooster | 0.11 | 0.53 | 0.30 | 0.16 | 0.14 | 0.17 | 0.23 |
| 17 | coast-forest | 0.13 | 0.36 | 0.30 | 0.26 | 0.28 | 0.32 | 0.20 |
| 21 | boy-sage | 0.04 | 0.51 | 0.24 | 0.16 | 0.14 | 0.32 | 0.32 |
| 25 | forest-graveyard | 0.07 | 0.55 | 0.30 | 0.33 | 0.20 | 0.22 | 0.25 |
| 29 | woodland-bird | 0.01 | 0.33 | 0.11 | 0.12 | 0.06 | 0.22 | 0.44 |
| 33 | woodland-hill | 0.15 | 0.59 | 0.49 | 0.29 | 0.25 | 0.32 | 0.23 |
| 37 | magician-ancient | 0.13 | 0.44 | 0.11 | 0.2 | 0.09 | 0.28 | 0.33 |
| 41 | sage-ancient | 0.28 | 0.43 | 0.11 | 0.09 | 0.05 | 0.32 | 0.35 |
| 47 | stove-furnace | 0.35 | 0.72 | 0.22 | 0.30 | 0.14 | 0.20 | 0.25 |
| 48 | magician-legends | 0.36 | 0.65 | 0.53 | 0.34 | 0.29 | 1.00 | 0.34 |
| 49 | mound-hill | 0.29 | 0.74 | 0.57 | 0.15 | 0.13 | 1.00 | 0.37 |
| 50 | cord-string | 0.47 | 0.68 | 0.55 | 0.49 | 0.34 | 0.80 | 0.42 |
| 51 | tumbler-glass | 0.14 | 0.65 | 0.52 | 0.28 | 0.25 | 0.80 | 0.40 |
| 52 | grin-smile | 0.49 | 0.49 | 0.60 | 0.32 | 0.33 | 1.00 | 0.46 |
| 53 | slave-former | 0.48 | 0.39 | 0.50 | 0.44 | 0.46 | 0.47 | 0.45 |
| 54 | voyage-make | 0.36 | 0.52 | 0.43 | 0.41 | 0.26 | 0.80 | 0.48 |
| 55 | autograph-signature | 0.41 | 0.55 | 0.43 | 0.19 | 0.33 | 0.80 | 0.41 |

| No | Corresponding word pairs | Mean | Li 2006 | Tsatsaronis 2010 | Islam 2008 | Ho 2010 | Croft 2013 | Islam 2012 |
|---|---|---|---|---|---|---|---|---|
| 56 | coast-shores | 0.59 | 0.76 | 0.93 | 0.47 | 0.49 | 0.80 | 0.42 |
| 57 | woodland-forest | 0.63 | 0.7 | 0.61 | 0.26 | 0.34 | 1.00 | 0.47 |
| 58 | implement-tool | 0.59 | 0.75 | 0.74 | 0.51 | 0.56 | 0.80 | 0.67 |
| 59 | cock-rooster | 0.86 | 1.00 | 1.00 | 0.94 | 0.87 | 1.00 | 0.53 |
| 60 | boy-lad | 0.58 | 0.66 | 0.93 | 0.6 | 0.57 | 0.80 | 0.62 |
| 61 | pillow-cushion | 0.52 | 0.66 | 0.35 | 0.29 | 0.26 | 0.80 | 0.49 |
| 62 | cemetery-graveyard | 0.77 | 0.73 | 0.73 | 0.51 | 0.59 | 1.00 | 0.48 |
| 63 | automobile-car | 0.56 | 0.64 | 0.79 | 0.52 | 0.38 | 1.00 | 0.64 |
| 64 | midday-noon | 0.96 | 1.00 | 0.93 | 0.93 | 0.86 | 1.00 | 0.87 |
| 65 | gem-jewel | 0.65 | 0.83 | 0.82 | 0.65 | 0.61 | 1.00 | 0.88 |